*Article*

# Monte Carlo study elucidates the type 1/type 2 choice in apoptotic death signaling in normal and cancer cells

Subhadip Raychaudhuri [1,*] and Somkanya Das [2,3]

[1] Department of Chemistry, University of California Davis; E-Mails: raychaudhuri@ucdavis.edu, subraychaudhuri@gmail.com.

[2] Genome Center, [3] Department of Biomedical Engineering, University of California Davis; E-Mails: scdas@ucdavis.edu.

**\*** Author to whom correspondence should be addressed; E-Mail: subraychaudhuri@gmail.com; Tel.: +1-530-574-6868.

**Abstract:** Apoptotic cell death is coordinated through two distinct (type 1 and type 2) intracellular signaling pathways. How the type 1/type 2 choice is made remains a fundamental problem in the biology of apoptosis and has implications for apoptosis related diseases and therapy. We study the problem of type 1/type 2 choice *in silico* utilizing a kinetic Monte Carlo model of cell death signaling. Our results show that the type 1/type 2 choice is linked to deterministic versus stochastic cell death activation, elucidating a unique regulatory control of the apoptotic pathways. Consistent with previous findings, our results indicate that caspase 8 activation level is a key regulator of the choice between deterministic type 1 and stochastic type 2 pathways, irrespective of cell types. Expression levels of signaling molecules downstream also regulate the type 1/type 2 choice. A simplified model of DISC clustering elucidates the mechanism of increased active caspase 8 generation, and type 1 activation, in cancer cells having increased sensitivity to death receptor activation. We demonstrate that rapid deterministic activation of the type 1 pathway can selectively target those cancer cells, especially if XIAP is also inhibited; while inherent cell-to-cell variability would allow normal cells stay protected.

## 1. Introduction

Apoptosis is the mode of cell death under a wide variety of cellular and physiological situations ranging from developmental regulations to tissue homeostasis. Dysregulation of apoptotic cell death has been implicated in a large number of diseases that includes degenerative disorders, cancer and autoimmune diseases. It has long been known that apoptosis is coordinated through two distinct pathways: the type 1 (extrinsic) pathway that does not involve mitochondria and the type 2 (intrinsic) pathway that requires mitochondrial activation. The extrinsic pathway is also called the death receptor mediated pathway as death receptor activation frequently induces apoptosis through this pathway. For a set of type 2 specific apoptotic stimuli (such as under DNA damaged conditions) only the type 2 pathway is selectively activated. However, for a wide range of cellular and physiological situations apoptosis is triggered by the activation of caspase 8 (initiator caspase) and thus could involve both pathways of apoptosis. Any types of death receptor (Fas, TNFR, DR4, DR5) mediated apoptosis fall into this category. A long-standing question in the biology of apoptosis is how the two pathways (type 1 and type 2) are differentially activated [1, 2, 3, 4, 5]. Some of the initial studies indicated that caspase 8 activation level coordinate the activation through those two pathways [1, 6]. Strong caspase 8 activation was implicated in the type 1 choice whereas weak activation of caspase 8 was thought to choose the type 2 pathway for signal amplification. It was also thought that caspase 8 activation is cell type specific and cells were labeled as either type 1 or type 2 depending on their choice of the activation pathway. We have utilized Monte Carlo simulations to elucidate that for low caspase 8 concentration the activation is dominated by the type 2 pathway with slow all-or-none type activation and large cell-to-cell variability, while for large caspase 8 concentration the type 1 pathway is activated in a rapid deterministic manner [7]. These findings make the problem of type 1/type 2 choice even more intriguing as it becomes linked to deterministic/stochastic choice in apoptosis activation. The stochastic type 2 to deterministic type 1 transition was shown to be a robust feature of apoptosis signaling irrespective of cell types [8]. A recent experiment indicated that the expression of death receptors coordinate the type 1/type 2 choice and both pathways can be activated in any cell types by regulating the expression of death receptors [5]. Variation in the expression of death receptors presumably regulates the activation level of caspase 8 and thereby determines between type 1 and type 2 pathways. However, the mechanisms for death receptor mediated caspase 8 activation have not been clearly elucidated. It is also not clear how expression levels of various downstream signaling molecules in the two apoptotic pathways affect the type 1/type 2 choice.

Previous studies in apoptotic cell death and apoptosis related diseases were dominated by population level measurements. More recent studies, however, indicate the presence of large cell-to-cell stochastic variability in the type 2 pathway of apoptosis [7, 9, 10, 11]. It has been shown that cell-to-cell variability in type 2 apoptotic activation is characterized by slow progression but eventual all-or-none type activation for single cells. Inter-cellular variability in the expression levels of apoptotic proteins, such as due to stochastic gene regulations, may introduce variability in the slow progression of apoptotic activation. This type of variability in apoptosis progression, coupled with rapid reaction events of cytochrome c release or caspase 9 activation, can lead to all-or-none type behavior. However, even when all the cellular parameters remain identical, inherent variability in apoptotic signaling reactions (due to small number of molecules or low-probability reactions) is capable of generating large cell-to-cell stochastic variability [7, 8, 9, 10, 12]. Both types of cell-to-cell variability can have important ramifications for diseases characterized by dysregulated apoptosis. In cancer, for example, cellular variation in anti-apoptotic protein levels (such as the Bcl2 concentration) as well as effective low number of molecules generated by over-expressed apoptotic inhibitors contribute to large cell-to-cell variability in apoptotic activation [10, 13, 14].

The programmed cell death mechanism of apoptosis utilizes a complex intracellular signaling network to carry out cellular demise in a controlled manner. In addition, phagocytic clearance of apoptotic bodies does not usually generate inflammatory conditions [15]. Targeting the apoptotic pathway is emerging as a new frontier in therapies of many of the apoptosis related diseases such as cancer. However, large cell-to-cell variability, including inherent variability, can lead to fractional cell killing under chemotherapy and thus pose a challenge for developing cancer therapies that target the apoptotic pathway. Incomplete activation of the type 2 pathway may lead to generation of more resistant phenotype. Therefore, it is important to search for strategies that would eliminate cell-to-cell variability in cancer cell apoptosis [13, 14, 16]. In addition, such strategies need to selectively target cancer cells. In this context, it is important to note that the expression levels of pro- and anti-apoptotic proteins vary significantly between normal and cancer cells [12, 17, 18, 19, 20]. In some cancer cells, which are equipped with increased levels of pro-apoptotic proteins Bid and Bax, it might be possible to reduce cell-to-cell variability by inhibiting Bcl2 like anti-apoptotic proteins and altering the Bcl2 to Bax ratio. Stochastic variability in Bax activation can even be abolished if a stochastic-to-deterministic transition is achieved [12]. However, in cancer cells for which the Bcl2 to Bax ratio has reached a very high level, alternative strategies such as switching the activation from type 2 to type 1 may provide an option [7, 20]. Thus the question of type 1/type 2 choice is highly relevant for cancer cell apoptosis and cancer therapy.

In this work, we utilize Monte Carlo simulations to explore the systems level regulatory mechanisms of the type 1/type 2 choice. Expression levels of various proteins involved in the regulation of type 1 and type 2 pathways were varied in isolation or in combination, and, activation of type 1 and type 2 pathways were measured under those conditions. Results obtained from our *in silico* studies indicate that the type 1/type 2 choice is regulated at a systems level by coordinated expression levels of signaling molecules in apoptotic pathways. Concentration of active caspase 8 (initiator caspase) emerges as a key regulator of the type 1/type 2 choice, consistent with previous studies [1, 6, 7]. Our results also indicate a key role of the apoptotic inhibitor XIAP, as well as the XIAP to Smac ratio, in the type 1/type 2 choice and systems level regulation of apoptosis [3, 4]. The formation rate of apoptosome is also shown to be important as its slow formation is a key rate limiting step in the type 2 pathway. In cancer cells, altered expression of various pro- and anti- apoptotic signaling proteins impact the type 1/type 2 choice. We demonstrate that increased sensitivity to death receptor activation in certain cancer cells can allow selective targeting of those cells (such as by death ligands) resulting in selective activation of caspase 8 only in those cells. XIAP inhibition in such death ligand treated cancer cells may result in a mixed type1-type 2 (or type 2) to type 1 transition in apoptotic activation and thus elimination of large cell-to-cell stochastic variability.

## 2. Experimental Section

A detailed computational study is carried out utilizing kinetic Monte Carlo (MC) simulations of pre- and post-mitochondrial signaling events [7]. A simplified network model of apoptosis signaling is studied that is triggered by active capsase 8 (Figure 1) [4]. Active caspase 8 initiates signaling through both type 1 and type 2 pathways. In the type 1 pathway, caspase 8 directly processes procaspase 3 to generate active caspase 3. In the type 2 pathway, caspase 8 cleaves Bid to an active form (tBid) that eventually lead to formation of active Bax dimers. Cytochrome c is released into the cytosol in an all-or-none manner when the number of active Bax dimers reaches a pre-assigned threshold value [21, 22]. Cytochrome c release leads to cytochrome c – Apaf binding and the subsequent formation of multi-molecular cyto c-Apaf-ATP complex apoptosome. Formation of the apoptosome complex is modeled in a simplified manner (by a cyto c-Apaf-Apaf-cyto c complex) where Apaf represents the Apaf-ATP complex. In Monte Carlo simulations, cytochrome c and Apaf molecules need to have spatial proximity before a binding reaction can occur between them, and such diffusion limitation introduces additional probability into the cytochrome c–Apaf binding reaction. Low probability of apoptosome formation generates stochastic variability in apoptotic activation, but once formed, induces rapid activation of downstream caspases 9 and 3 [7, 23]. Activation of caspase 3 (effector caspase) can be taken to be a downstream readout of apoptotic cell death signaling; MC simulations are carried out to measure the time-course of caspase 3 activation at a single cell level.

Figure 1. Schematic of the apoptotic death signaling network. Apoptosis is activated through two distinct pathways: type 1 (intrinsic) and type 2 (extrinsic). The type 1 – type 2 signaling loop is initiated by generation of active caspase 8 and ultimately converges on caspase 3/7 activation.

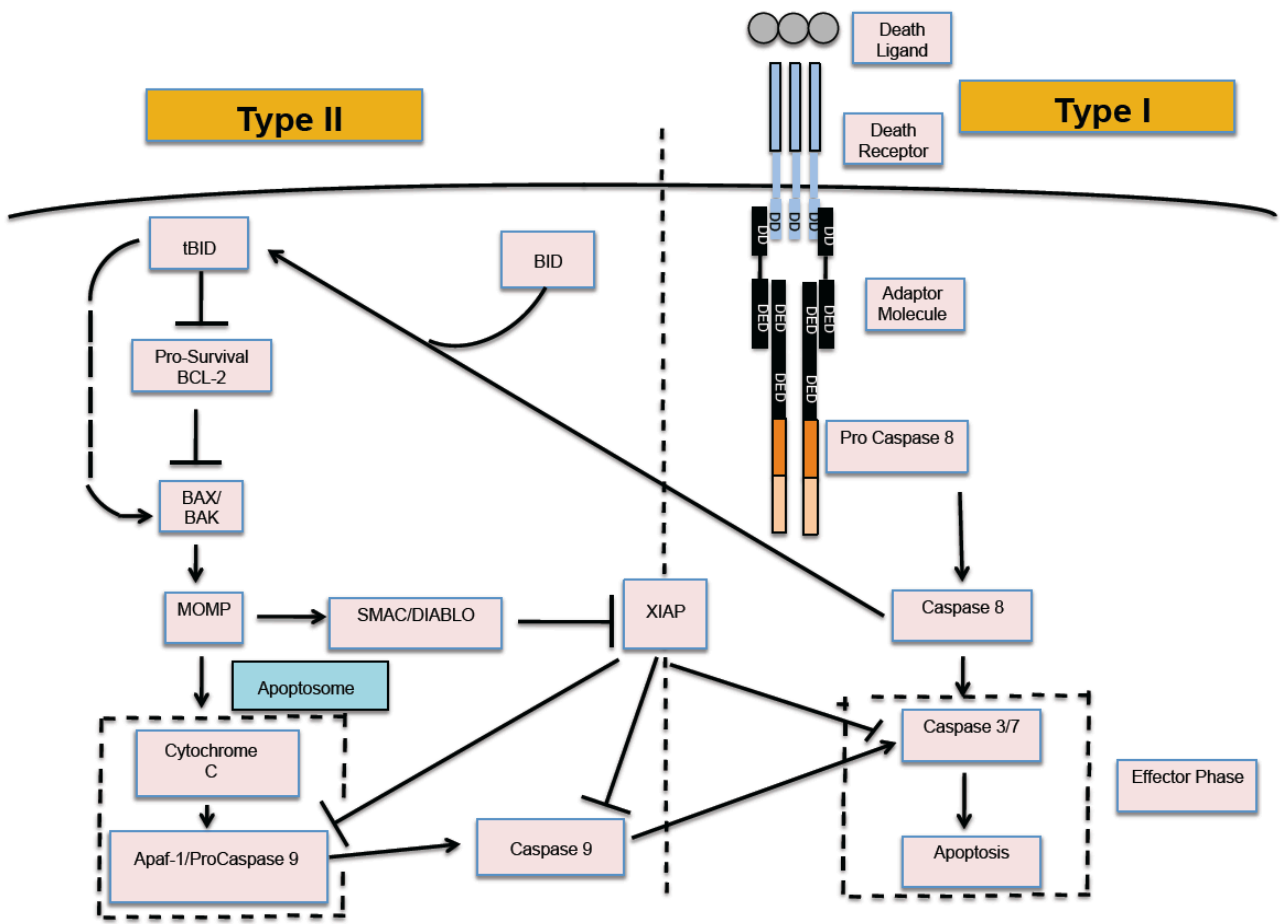

In our model, effects of functionally similar proteins are captured by a representative protein. Bcl-2 (B cell lymphoma protein 2) represents all the Bcl-2 family proteins (such as Bcl-2, Bcl-xL, Mcl-1) with similar anti-apoptotic properties. We do not explicitly simulate BH3 only sensitizers (such as Bad or Bik), whose effect can be simulated by varying Bcl-2 concentrations. Presence of multiple functionally similar proteins and their varied expression levels (depending on cell type) make the apoptotic pathway more complex than the present model can address. However, coarse-graining the pathway by representative proteins allows us to capture some of the essential biology of type 1/type 2 choice in apoptotic regulation.

The Monte Carlo approach is well suited to simulate some of the complexities of signaling reactions such as the spatial heterogeneity involved in signaling reactions. Bax molecules, for example, translocate to mitochondria upon activation. Caspase 8 activation is known to be mediated by clustering of adaptor proteins (such as FADD/TRADD) recruited to death receptor-ligand complexes. Procaspase 8 molecules are recruited to the clustered adaptor proteins to generate the assembly of DISC (death-inducing-signaling-complex) and generate active caspase 8 molecules through autoprocessing [24]. In the current study, a simplified model of DISC (death inducing signaling complex) formation is considered where adaptor molecules can cluster (to lower thermodynamic free energy) when they are bound to death receptor-ligand complex; we call this state (receptor-ligand complex bound) of the adaptor molecule an active state. The parameter that captures the reduced energy of two neighboring active adaptor molecules is denoted by $E_{DD}$ (is taken to be -2 $K_BT$ unless specified otherwise). DISC formation is incorporated into the simulation by a hybrid simulation scheme between kinetic Monte Carlo model of intracellular signaling with an explicit free energy based model for clustering of adaptor molecules [25, 26]. Effective probability parameters $P_{on}$ and $P_{off}$ are introduced that captures an adaptor molecule's switching between an active and an inactive state (to capture the effect of death ligand induction such as FAS/TRAIL binding). Simulations are carried out for various values of the parameters values of $P_{on}$ and $P_{off}$ (those presumably vary depending on the cell type and/or the receptor type).

Kinetic reaction rates (such as $k_{on}/k_{off}$) and molecular concentrations are obtained from values reported in the literature (such as [6]) and utilized in our previous work [7, 9, 10, 12, 13] (unless specified otherwise). At each Monte Carlo (MC) step one molecule is sampled (on average) twice to allow for one diffusion and one reaction move (only one move per molecule was allowed in previous implementations). Each MC step is chosen to be $10^{-4}$ s; a typical simulation is run until the concentration of unbound active caspase 3, the effector caspase, reaches its half-maximal value ~ 50 nM (simulation stops if it is not completed before a predefined number of MC steps). Certain threshold amount of caspase 3 activation might be sufficient for downstream PARP cleavage and that threshold should also regulate the type 1/type 2 choice. In this work, data normalized to its half-maximal value is reported for caspase 3 activation. Type 1/type 2 classification is also based on half-maximal activation of caspase 3. Simulation volume ($1.2 \times 1.2 \times 1.2$ μm$^3$) is chosen in such a manner that the number of molecules (for each molecular species) is equal to the nanomolar concentration. Controlled Monte Carlo experiments are carried out for specific parameter values (such as molecular concentrations). Each run of the simulation corresponds to activation at a single cell level.

## 3. Results and Discussion

*3.1 Increased active caspase 8 concentration switches the activation from stochastic type 2 to deterministic type 1*

We have previously shown that two apoptotic pathways are activated in a distinct manner [7, 8]. Apoptotic cell death can be activated in rapid deterministic manner through the type 1 pathway. In contrast, apoptotic activation through the type 2 pathway is usually slow with inherent cell-to-cell variability. How these two distinct pathways are activated remains an unresolved issue in the biology of apoptosis. Previous studies indicated the role of active caspase 8 as a key determinant of the type 1/type 2 choice [1, 2, 6, 7]. In Fig 1, the signaling network for apoptotic cell death is shown where the type 1 and type 2 pathways form a loop network that can be triggered by active caspase 8. Clearly, caspase 8 is expected to be a critical regulator of the type 1/type 2 choice and such a task is achieved by differential binding to its immediate binding partners procaspase 3 (type 1 pathway) and Bid (type 2 pathway). Capsase 8 binds to Bid with moderately high affinity ($k_A = k_{on} / k_{off} \sim 10^9$ M$^{-1}$) compared to its relatively weak affinity for procapsase 3 ($k_A = k_{on} / k_{off} \sim 1.67 \times 10^5$ M$^{-1}$). Thus for low concentrations of active caspase 8 (weak apoptotic stimuli) the type 2 pathway is preferentially activated (Figure 2a). Large caspase 8 concentration (strong apoptotic stimuli), on the other hand, is sufficient to directly activate procaspase 3 in the type 1 pathway (Figures 2b and 2c). Mixed type 1-type 2 behavior is also observed (where procaspase 3 is cleaved by both caspase 8 and caspase 9), especially for intermediate levels (~ 5 nM) of caspase 8 activation (Figure 2b). A large contribution to cell-to-cell variability in caspase 3 activation (Figure 2a) seems to originate from stochastic variability in apoptosome formation (post-mitochondrial signaling module) but remains sensitive to its rate of formation [9]. The type 2 pathway is designed to amplify an initially weak signal (to a strong all-or-none type activation) while large cell-to-cell variability generates heterogeneity in cellular response presumably as an adaptive strategy to a weak stimulus [7, 8].

In the limit of weak capsase 8 activation (~ 1 nM), a set of ODEs describing caspase 8 reactions to Bid and procaspase 3 can be utilized to estimate that caspase 8-Bid complex is ~ 10$^3$ fold more abundant than caspase 8- procaspase 3 complex (Appendix). It can also be shown that at early-time caspase 8-procaspase 3 complex concentration ~ $0.6 \times 10^{-3}$ [caspase 8], where [] denotes concentration of given molecule. Therefore, unless caspase 8 concentration is very high (~ 100 nm or higher), caspase 8 starts selectively activating the type 2 pathway by cleaving Bid to tBid. Even though the initial type 1/type 2 preference is determined by the amount of active caspase 8, the final type 1/type 2 choice should depend on how fast activation signal through the two pathways can lead to completion of caspase 3 activation.

Figure 2. Caspase 3 activation for three different caspase 8 concentrations: 1 nM (a), 5 nM (b), 10 nM (c). Data is shown for 10 representative single cell runs; each color corresponds to apoptosis activation for a single cell (Monte Carlo run). Type 2 activation is characterized by all-or-none type behavior with large cell-to-cell variability. Increasing concentration of caspase 8 results in increased type 1 activation and suppression of cell-to-cell variability.

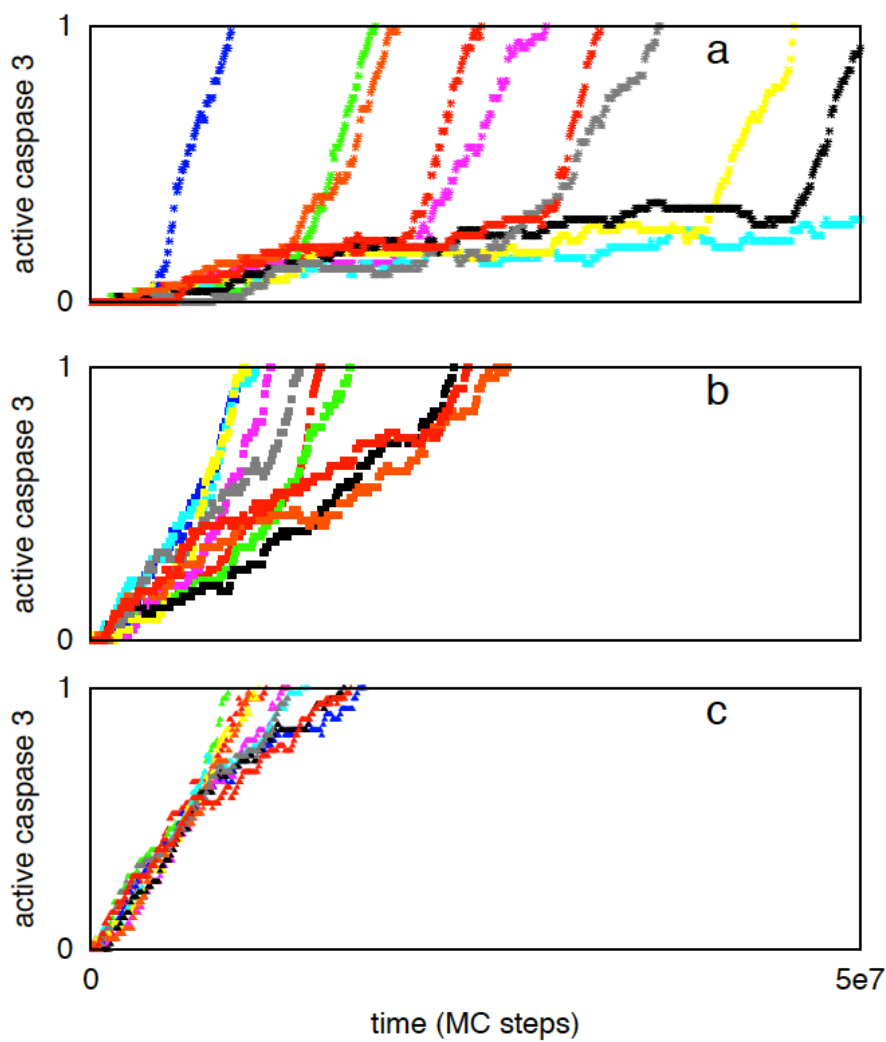

*3.2 Systems level regulatory mechanisms of type 1/type 2 choice*

Our previous studies elucidated mechanisms for slow activation and generation of cell-to-cell variability through the type 2 pathway: (1) effective small number of reactant molecules generated due to the action of inhibitory molecules such as Bcl2 and XIAP [10, 13] and (2) low probability reaction events such as the apoptosome formation [7]. These parameters are varied here to study their impact on the type 1/type 2 choice.

3.2.1 Effect of small (~ 3 fold) Bcl2 variation on cyto c release and type 1/type 2 choice

Small variations in Bcl2 levels can capture tissue specific variations as well as variations during various developmental stages [27]. Bcl2 over-expression (3 fold) resulted in inhibition of Bax activation and slightly delayed release of mitochondrial cytochrome c. Subsequent formation of XIAP-Smac and apoptosome complex were also delayed. For caspase 8 = 1 nM, the average time to apoptosome formation was higher when Bcl2 was 3 fold overexpressed compared with the same for regular Bcl2 expressions ($2.74 \times 10^7$ vs $2.39 \times 10^7$ MC steps). Slower apoptosome formation resulted in a slightly delayed type 2 activation and a small increase in type 1 activation. When active caspase 8 concentration was increased to 10 nM, type 1 pathway dominated the activation and Bcl2 inhibition no longer had a significant effect. Bcl2 is a known oncogene and large over-expression (> 10 fold) of Bcl2 is considered later in the context of cancer cells.

3.2.2 XIAP to Smac ratio is a key regulator of type 1/type 2 choice: inhibition of the type 1 activation by XIAP

XIAP is known to be a key regulator of the apoptotic activation as it binds and inhibits both caspase 9 (including procaspase 9) and caspase 3. Previous experimental studies indicated that under XIAP inhibition caspase 8 processing of procaspase 3 cannot proceed to its fully cleaved p17/p12 form, instead it remains in a partially processed p20/p12 form [3]. It was observed that XIAP could inhibit both p17/p12 and p20/p12 forms of active caspase 3. It was also shown that mitochondrial release of Smac, upon threshold Bax activation, can inhibit XIAP and thus modulates the type 1/type 2 choice in apoptosis activation [3]. We carried out simulations for various XIAP / Smac ratios [3, 28]. Initially XIAP concentration was kept constant at 30 nM while the Smac level was varied: 50 nM (XIAP to Smac ratio 0.6) and10 nm (XIAP to Smac ratio 3). When caspase 8 = 1 nM, activation is dominated by the type 2 pathway as observed in type 2 cells. Decreasing the Smac concentration from 50 to 10 nM resulted in increased XIAP inhibition of caspases and suppression of the initial type 1 activation (Fig 3a and 3b; Table S1). Similar behavior was observed when XIAP concentration was increased to 90 nM (Smac = 50 nM) (Fig 3c; Table S1). In this case, the type 2 activation was delayed (Fig 3c) until the XIAP-caspases 3 complexes reached sufficiently high concentration so that procaspase 9 molecules could be relieved from XIAP inhibition. For XIAP = 60 nM and Smac = 50 nM, removing Bid from the system resulted in weak type 1 activation (Fig S1a). However, when XIAP was inhibited (in Bid deficient cells), slow type 1 activation was observed (Fig S1 b and c) consistent with earlier experimental studies of Fas ligand induced apoptosis in Bid$^{-/-}$XIAP$^{-/-}$ mice hepatocytes (type 2 cells) [4]. When the initial XIAP concentration is set to zero in presence of Bid, more rapid apoptotic activation was observed and for a few cells half-maximal activation was nearly completed through the type 1 pathway (Fig S2). Type 1 activation became increasingly dominant for increased active caspase 8 (~ 2-3 nM) (Fig S2). Even in presence of significant amount of XIAP, when active caspase 8 concentration was increased to ~ 5 nM (or higher) significant type 1 activation was observed along with the type 2 signaling (Fig 3d and 3e; Fig S3). Mitochondrial Smac release inhibited XIAP and assisted type 1 activation to proceed. Such a mode of apoptotic activation can be labeled as type 2 assisted type 1 activation. Eventually, caspase 9 was activated and subsequent activation of caspase 3 progressed through both type 1 and type 2 pathways (mixed type 1-type 2 behavior) (Fig S3). When XIAP was highly expressed (~ 90 nM), type 1 activation was observed to proceed only after XIAP was inhibited by mitochondrial Smac release (Fig 3f) [3]. The effect of type 1 assisted type 2 activation was diminished when Smac release was suppressed in a significant manner. In addition, reduced Smac concentration resulted in slower rate of caspase 9 activation (as more XIAP becomes available to inhibit procaspase 9) and slower apoptosis. Therefore, apoptotic inhibitor protein XIAP (also the XIAP to Smac ratio) emerges as a key determinant of the type 1/type 2 choice in apoptotic activation. The impact of XIAP inhibition in type 1/type 2 choice can be explained by noting that XIAP inhibits several apoptotic molecules in both type 1 and type 2 pathways; in addition, it inhibits the effector caspases (such as caspase 3) that close the type 1 – type 2 loop network.

Figure 3. Single cell caspase 3 activation for different values of XIAP to Smac ratio and active caspase 8 levels. Two distinct active caspase 8 concentrations are studied: 1 nM (left panel: a,b,c) and 10 nM (right panel d,e,f). The following XIAP and Smac levels are simulated: XIAP = 30 nM, Smac = 50 nM (a and d), XIAP = 30 nM, Smac = 10 nM (b and e), XIAP = 90 nM, Smac = 50 nM (c and f). Data is shown for 5 representative single cells in each of the above cases; each color corresponds to apoptosis activation for a single cell (Monte Carlo run).

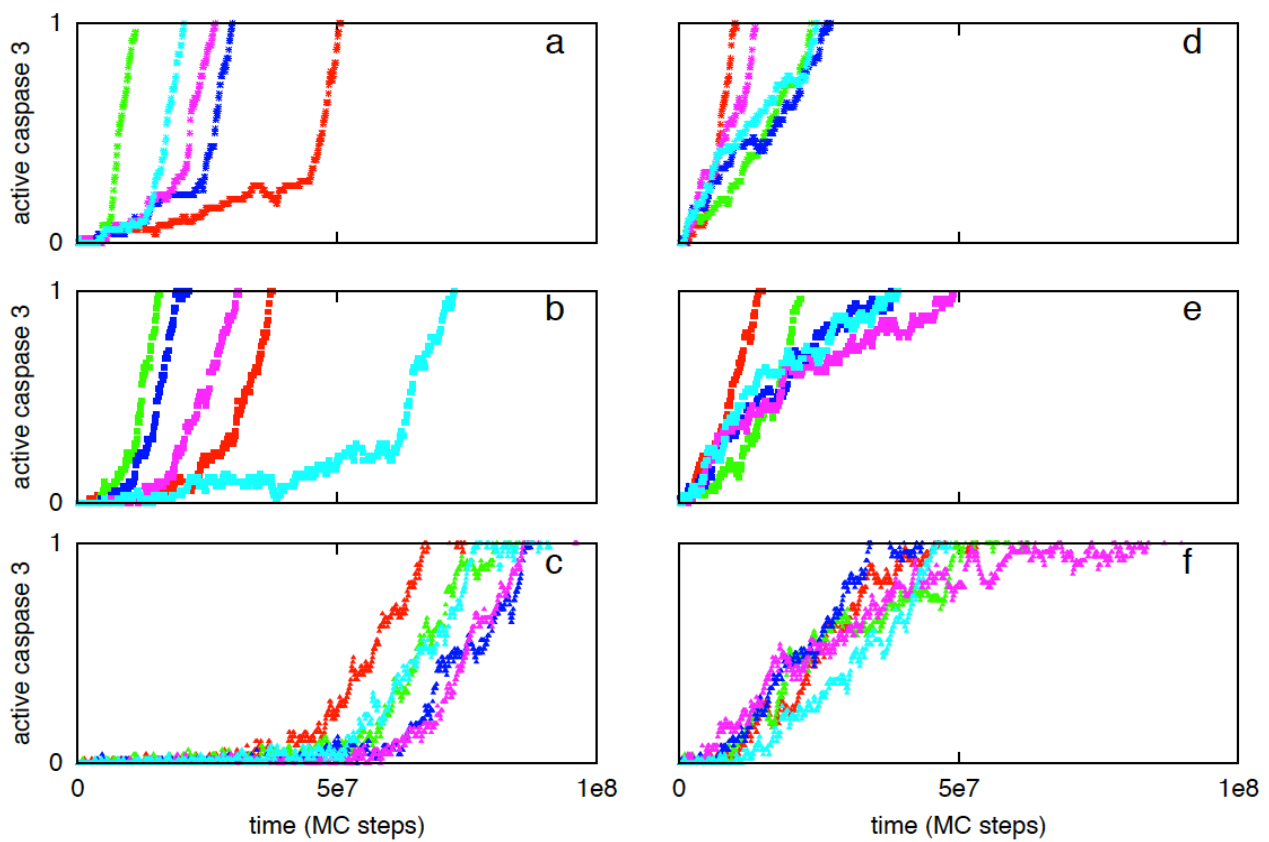

3.2.3 Increased rate of apoptosome formation favors the type 2 pathway

To study the effect of slow apoptosome formation on type 1/type 2 choice we varied the Apaf concentration (~100 nM in ref [6] and ~20 nM in ref [29]; [30, 31]) and the kinetic reaction rate $k_{on}$ for the cytochrome c – Apaf association reaction ($k_{on}$ may vary under pH change [32]). For low concentrations of caspase 8 (=1 nM) and when Apaf = 100 nM, activation is dominated by the type 2 pathway (Figure 2a). When Apaf concentration was reduced to 20 nM, slow type 1 activation frequently replaced type 2 activation (Figure S4a). For low Apaf concentrations, low-probability apoptosome formation becomes very slow allowing type 1 activation to complete before caspase 9 is activated (Figure S4). When the low probability rate constant of cytochrome c and Apaf binding was assumed 2 fold higher, formation of apoptosomes was faster leading to increased type 2 activation but diminished cell-to-cell variability. In contrast, when the cyto c-Apaf binding rate was taken 2 fold lower, increased type 1 signaling was observed due to slower rate of formation of apoptosomes. The average times to form the first apoptosome are provided in Table S2 (supplemental material). For a large (~10 fold or higher) increase in the cytochrome c-apaf association constant, cell-to-cell variability generated through the post-mitochondrial signaling module is significantly decreased.

3.3 *Robustness of type 1/type 2 choice in apoptosis regulation*

Previously we utilized a minimal signaling network (Fig S5) to demonstrate that stochastic type 2 to deterministic type 1 transition in apoptosis activation can be achieved irrespective of cells types (provided a few simple conditions are satisfied) [8]. The same minimal network is studied here to show that some of the systems levels regulatory mechanisms of the type 1/type 2 choice, as observed in Monte Carlo simulations of the present work, are robust features of apoptosis signaling (supplemental material). Specifically, we varied the following variables in the minimal network: (i) the concentration of the signaling molecule that opens the type 1/type 2 loop and (a measure of strength of the apoptotic stimulus) (ii) the rate constant for the slow step in the type 2 pathway. The signaling molecule that opens the type 1/type 2 loop in the minimal network captures the effect of active caspase 8 in the full apoptotic network. Increasing (decreasing) its concentration resulted in increased (decreased) activation of the type 1 pathway (Fig S6); result was shown to be robust over 2 fold variations in concentrations of other signaling molecules. Increasing (decreasing) the rate constant for the slow step in the type 2 pathway, which could capture variation in the formation rate of apoptosome, enhanced (diminished) the type 2 activation (Fig S7). These results are consistent with findings of the in silico experiments of the apoptotic death pathway and elucidate some of the robust regulatory mechanisms of type 1/type 2 choice in apoptosis signaling. The precise concentrations of different molecules (in the full apoptotic network) at which the type 1/type 2 transition occurs should depend on cell type specific features and can be better predicted as more accurate quantitative information (such as regarding molecular concentrations and rate constants) becomes available.

3.4 *Regulation of type 1/type 2 choice in cancer cells having over-expressed anti-apoptotic proteins Bcl2 and XIAP*

A hallmark of cancer is genomic instability and aberrant expression of oncogenes and proteins. Expression levels of signaling proteins in the apoptotic death pathway are particularly altered, including the level of well known oncogenic proteins such as Bcl2, resulting in dysregulated death signaling and inhibition of apoptosis. Previous studies elucidated the mechanism of Bcl-2 and XIAP inhibition of type 2 apoptosis in cancer cells [13]. It was shown that increased expression of Bcl2 and/or XIAP enhances cell-to-cell stochastic variability and time-to-death in type 2 activation, imparting apoptosis resistance to cancer cells. In this work, we focus on the regulation of type 1/type 2 choice in cancer cells that are equipped with increased levels of Bcl2 and XIAP. We vary the concentrations of Bcl2 and XIAP and study type 1/type 2 control of apoptosis. A 2 fold over-expression for Bid and Bax molecule is also assumed [12].

Figure 4. Time course of caspase 3 activation in cancer cells having over-expressed Bcl2 and XIAP. Bcl2 over-expression is taken as ~15 fold while XIAP expression is varied: normal (a and d), 3 fold (b and e), 5 fold (c and f). Two different caspase 8 concentrations are used: 1 nM (left panel: a,b,c) and 10 nM (right panel: d,e,f). Data is shown for 5 representative single cells in each of the above cases; each color corresponds to apoptosis activation for a single cell (Monte Carlo run).

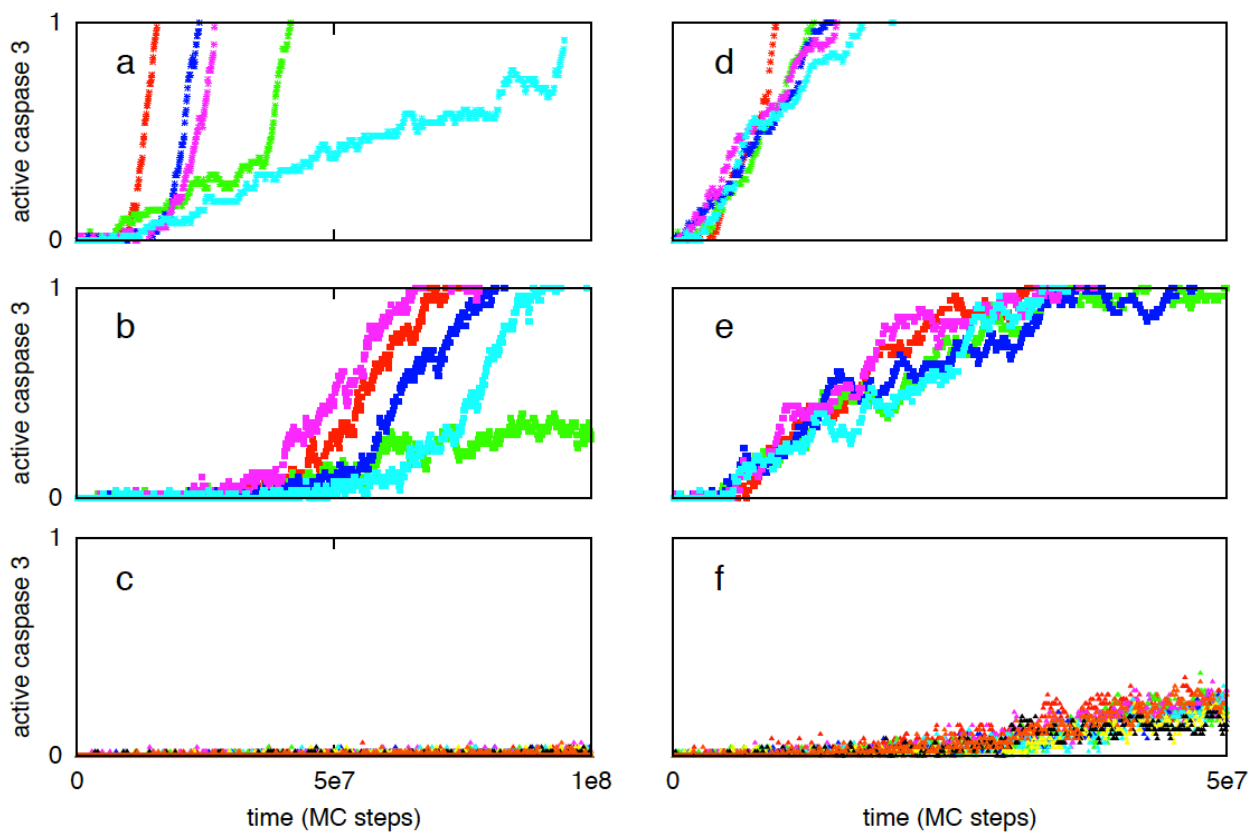

Bcl-2 over-expression resulted in slower cytochrome c / Smac release with increased cell-to-cell stochastic variability, consistent with observations of previous studies [10, 12]. When Caspase 8 = 1 nM and XIAP = 30 nM, 15 fold Bcl2 over-expression allowed initial slow type 1 activation to proceed longer but eventually switched to all-or-none type 2 activation (Figure 4a). Cell-to-cell variability in caspase 3 activation (Figure 4a) results from stochastic variability in (i) Cytochrome c/Smac release (pre-mitochondrial signaling module) and (ii) apoptosome formation (post-mitochondrial signaling module) (Table S3). When XIAP is 3 fold over-expressed strong suppression of the initial type 1 activation was observed and caspase 9 activation precedes caspase 3 activation (Figure 4b). For high XIAP over-expression (~ 5 fold of normal 30 nM), apoptotic activation is strongly suppressed (Figure 4c) providing a mechanism for resistant cancer cells. Certain cancer cells are known to be chemosensitive to death ligand induction presumably by having increased sensitivity to death receptor activation [19]. In such cancer cells it is expected that increased death receptor activation would result in increased amount of active caspase 8 generation. When caspase 8 = 10 nM but XIAP level remains normal (30 nM), rapid deterministic activation of the type 1 pathway is observed (irrespective of Bcl2 over-expression levels) (15 fold Bcl2 in Figure 4d). Therefore, achieving such a caspase 8 activation level (~ 10 nM) combined with lower XIAP level (~ 30 nM) can be an optimal strategy for targeting certain cancer cells. Increased XIAP expression, however, inhibits the type 1 pathway of apoptosis. When XIAP is 3 fold over-expressed along with 15 fold over-expression of Bcl2, type 1 activation is suppressed until Smac is released from mitochondria. Smac binding to XIAP allows caspase 3 activation to proceed and subsequent activation of caspase 9 induces additional caspase 3 activation through the type 2 pathway (Figure 4e). Hence, a type 2 assisted type 1 mode of activation followed by a mixed type1-type2 activation was observed when XIAP was moderately over-expressed (~ 3 fold). When Bcl2 over-expression was taken to be 50 fold (keeping XIAP ~ 3 fold), Cyto c / Smac release became slower with increased cell-to-cell stochastic variability. As a result, caspase 3 activation showed large cell-to-cell stochastic variability with all-or-none type behavior (Figure S8). Increasing the XIAP over-expression level to 5 fold led to strong XIAP mediated inhibition of caspase 3 activation and weak type 1 activation (Figure S8 c and f). Additional interactions such as caspase 6 mediated activation of caspase 8 (positive feedback) [3] and/or XIAP degradation of active caspase 3 can further impact the type 1/type 2 choice in a cell type dependent manner.

3.5 *Cancer cell apoptosis with heightened sensitivity to death receptor activation: a mechanism for selective targeting of cancer cells*

Targeting the apoptotic pathway can be an effective way to eliminate cancer cells in a controlled manner. However, one needs to find a strategy that will selectively target cancer cells leaving normal cells protected. It is known that in certain cancer cells significant expression of death receptors (such as Fas/TNFR/TRAIL (DR4/DR5)) provides an opportunity to selectively target cancer cells [19, 33, 34]. In addition, normal cells might have protection from larger expression of decoy receptors (compared to that in cancer cells) [35, 36] resulting in effective reduced expression of death receptors. The specificity for cancer cell apoptosis can be induced (or enhanced) by additional mechanisms such as gene therapeutic approaches [37]. The specific choice of death ligand (or agonist antibody) depends on a variety of factors such as whether the targeted death receptor generates type 1 or type 2 activation in other types of normal cells [19]. Even though mechanisms of activation and downstream effects may vary depending on the specific death receptor under consideration, the mechanism of death receptor mediated caspase 8 activation (from procaspase 8) seems to be similar in many of those cases and involves (i) recruitment of adaptor molecules (to ligand bound death receptors) containing both death domain and death effector domain (such as FADD) and (ii) formation and clustering of DISC [24]. To capture selective killing of cancer cells we study apoptosis mediated by death receptor induced caspase 8 activation. The mechanism of death receptor clustering (such as by death ligand induction) is modeled in a simplified manner by DISC clustering (see Experimental Section). To capture the effect of death ligand induction (such as by TRAIL), we carry out simulations for the various $P_{on}$ and $P_{off}$ values. Even though some clustering was started to be seen for $P_{off}$ values ~ $10^{-2}$, significant clustering was observed when $P_{off} = 10^{-3}$ (such as in cancer cells). DISC clustering is shown in the supplemental material (Fig S9). Corresponding caspase 8 activation are shown in Figure 5. Lower values for the DISC free energy parameter $E_{DD}$ resulted in increased caspase 8 activation (Fig 5). The average time-to-death ($T_d$) decreased with increased DISC clustering and caspase 8 activation ($T_d = 4.46 \times 10^7$ MC steps for $E_{dd} = 0$; $T_d = 2.95 \times 10^7$ MC steps for $E_{dd} = -K_BT$; $T_d = 2.93 \times 10^7$ MC steps for $E_{dd} = -2K_BT$). Slower generation of active caspase 8 also provides a mechanism for increased cell-to-cell stochastic variability. For normal cells, reduced expression of death receptors and the inhibitory effect of decoy receptors is captured by lower $P_{on}$ values of $10^{-4}$ (16% cell death at t = $10^8$ MC steps) and $10^{-5}$ (1.6% cell death at t = $10^8$ MC steps) and $P_{off} = 1$.

Figure 5. Fraction of caspase 8 activation for 3 different values of DISC clustering energy: (a) $E_{dd} = 0$, (b) $E_{dd} = -K_BT$, (c) $E_{dd} = -2K_BT$. Pon = 1 and Poff = $10^{-3}$ are assumed. We simulated cancer cells having over-expressed Bid (2 fold), Bax (2 fold), Bcl2 (15 fold) and XIAP (3 fold). Data analyzed from 64 single cells (MC runs) are shown for three different time instants: $5 \times 10^6$, $10^7$ and $3 \times 10^7$ MC steps. Clustering of DISC (when $E_{dd} = -K_BT$ or $-2K_BT$) resulted in increased active caspase 8 generation and subsequent activation of effector caspases.

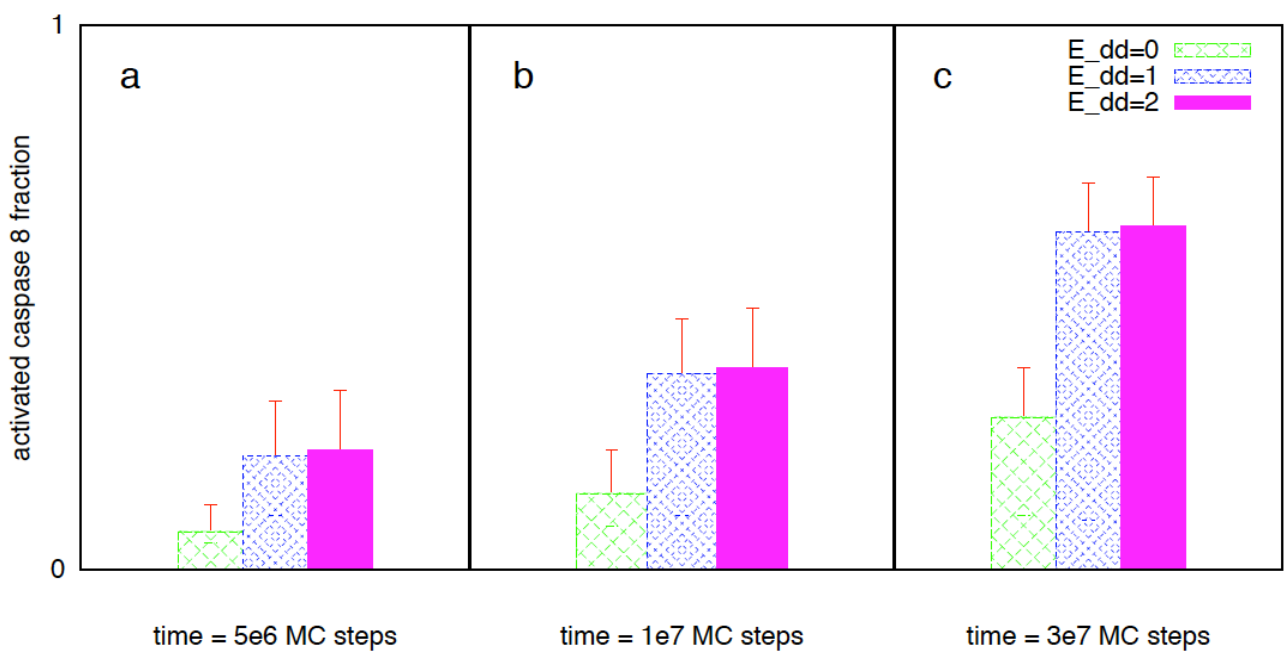

The mechanism of death receptor clustering (such as by death ligand induction) and subsequent caspase 8 activation is complex. Formation of closed oligomers (such as of Fas-FADD [38]) within the DISC and/or altered partitioning of membrane proximal molecules in lipid rafts (upon ligand induction) may lead to a stronger effect of DISC formation on caspase 8 activation [39, 40]. A detailed model that will explicitly simulate death receptor-ligand binding and lipid mediated interactions is currently under development. The strength of an external apoptotic stimulus, such as concentrations of a death ligand, is expected to affect death receptor activation and the type 1/type 2 choice, but the mechanism remains to be explored. Both apoptotic activation as well as protection (from apoptosis) due to over-expressed Bcl-$x_L$ have been shown to correlate with death ligand concentrations [3, 5]. Studies in neural cells indicated activation of both pathways when treated with Aβ-oligomers (type 2 activation) and Aβ-aggregates (type 1 activation) [41]. In addition to its key role in apoptosis, mechanisms of death receptor activation and caspase 8 generation have implications for other modes of cell death such as in necroptosis [42, 43].

Cancer drugs that target the apoptotic pathway by perturbing the membrane lipid environment have attracted recent attention [44, 45]. The mechanisms of death receptor activation may also have implications for the cancer stem cell model. Recent studies in severely immunocompromised NOD/SCID IL-2rg$^{-/-}$ mice found significantly increased fraction of tumorigenic cells (melanoma stem cells) when compared with the same in NOD/SCID mice [46]. Cellular and molecular basis of such increased fraction of cancer stem cells in NOD/SCID IL-2rg$^{-/-}$ mice is yet to be elucidated. However, a role of NK cell mediated tumor cell killing can be postulated as NOD / SCID IL-2rg$^{-/-}$ mice are known to be depleted of NK cells (due to the null Interleuken-2 receptor gamma chain mutation). NK cells are equipped with death ligands that are known to activate the death receptors DR4 and DR5 on target cells [47]. Therefore, NK cell mediated apoptotic activation may provide a mechanism for increased killing of tumor cells transplanted in NOD/SCID mice.

*3.6 Combined death ligand induction and XIAP inhibition can be an optimal strategy to kill cancer cells: maximizing specificity and minimizing cell-to-cell stochastic variability*

Highly over-expressed anti-apoptotic proteins allow cancer cells to resist apoptotic cell death. Increased inherent cell-to-cell variability, in apoptotic activation of cancer cells equipped with over-expressed anti-apoptotic proteins, can be a mechanism for the resistant phenotype of cancer cells. Fractional killing of cancer cells (under chemotherapy) need to be addressed as survived cells may acquire more resistant phenotype. In a recent work we have shown that over-expression of certain pro-apoptotic proteins, along with over-expressed apoptotic inhibitors, can allow a stochastic-to-deterministic transition in the type 2 pathway (by simply inhibiting Bcl2 like proteins in the type 2 pathway) [12]. While such a strategy might be effective for many types of cancer cells, there are cancer cells that will remain resistant due to very high Bcl2 to Bax ratio. For cancer cells that are resistant to Bcl2 inhibitor chemotherapy, switching the activation from type 2 to type 1 can be a potential option for inducing rapid deterministic apoptosis. Such a strategy is demonstrated here for cancer cell types having heightened sensitivity to death receptor activation but also high Bcl2 to Bax ratio ( ~ 25). Overexpressions for the following signaling molecules are assumed: Bid (2 fold), Bax (2 fold), Bcl2 (50 fold) and XIAP (5 fold). We perform in silico experiments for the following chemotherapeutic targeting of cancer cells: death ligand induction and various degree of Bcl2 and/or XIAP inhibition. Increased sensitivity to death receptor activation (upon death ligand induction) was captured by our simplified model of DISC clustering ($P_{on} = 1$, $P_{off} = 10^{-3}$, $E_{dd} = -2\ K_BT$).

Figure 6. Fraction of caspase 3 activation for 4 different chemotherapeutic strategies involving DISC formation and one of the following: (i) Bcl2 inhibition (50 fold to 15 fold), (ii) stronger Bcl2 inhibition (50 fold to 5 fold), (iii) Bcl2 (50 fold to 15 fold) and XIAP inhibition (5 fold to 3 fold) and (iv) strong XIAP inhibition (5 fold to normal) Cancer cells are assumed to have high over-expression levels for both Bcl2 (~ 50 fold) and XIAP (~ 5 fold). The following parameters are utilized for the DISC clustering: Pon = 1, Poff = $10^{-3}$, $E_{dd}$ = -2 $K_BT$. Data analyzed from 64 single cells (MC runs) are shown for three different time instants: $2.5 \times 10^7$, $5 \times 10^7$ and $1 \times 10^8$ MC steps.

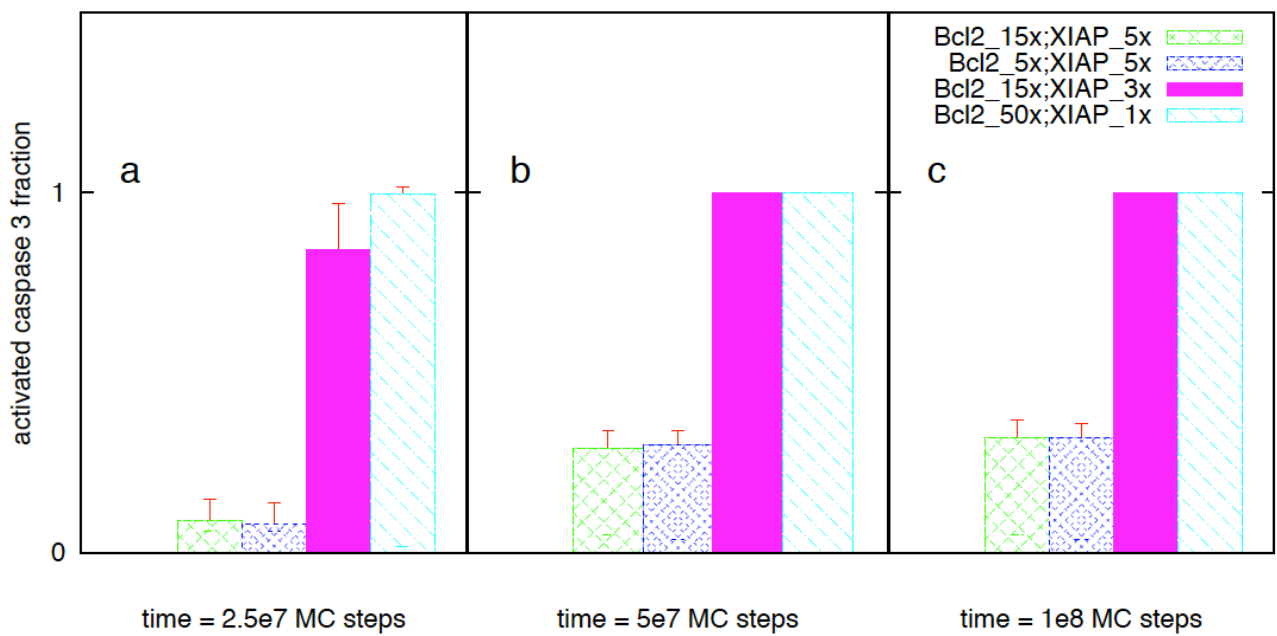

DISC formation resulted in significant caspase 8 activation (~ 10 nM). However, XIAP over-expression led to caspase 3 inhibition and strong suppression of apoptotic activation. We observed some capsase 3 activation at long times ($5 \times 10^8$ MC steps) that can be characterized as mixed type 1 and type 2 activation. Inhibiting Bcl2 (50 fold over-expression reduced to 15 fold or 5 fold) did not alter the strong suppression in apoptotic activation (Fig 6). Combined Bcl2 (50 fold to 15 fold) and XIAP inhibition (5 fold to 3 fold) resulted in significantly increased apoptotic activation involving both type 1 and type 2 pathways (Fig 4e and 6). However, when XIAP was strongly inhibited type 1 activation could proceed in a fast deterministic manner (Fig S8 d and Fig 6). Such a strategy might also allow to selectively target cancer stem cells having heightened sensitivity to death receptor activation.

Heightened sensitivity to death receptor activation in cancer cells can be accompanied by simultaneous increase in anti-apoptotic factors resulting in a dichotomous membrane proximal signaling module. Therefore, in spite of the heightened sensitivity to death receptor activation in cancer cells (and possibly also in cancer stem-like cells), inhibitory effects of various anti-apoptotic factors may only allow low capsase 8 activation leading to type 2 activation (Fig 4a and Fig S8a). The inhibitory effects can be potentiated by membrane-proximal factors such as increased expression of cFLIP [48], increased amount of decoy receptors [49] or altered partitioning of death receptors in membrane domains. However, death ligand induced weak apoptotic activation through the type 2 pathway can be converted to rapid deterministic activation through the type 1 pathway by eliminating the increase in anti-apoptotic factors (in the membrane proximal signaling module) and/or enhancing death receptor density [5]. It has been shown that cFLIP inhibition can render cancer cells susceptible to death receptor mediated apoptotic death [48]. Expression of death receptors can be increased by inducing transcription [37], increasing transport to membranes, or even by freeing them from other binding partners [50, 51]. If XIAP is also over-expressed, then simultaneous activation of death receptors and inhibition of XIAP will lead to rapid activation of the type 1 pathway (Fig 6). In normal cells lack of sensitivity to death receptor activation will lead to protection. Slight over-expression of death receptors in some normal cells might initiate type 2 activation, but slow activation with large inherent cell-to-cell variability will eventually protect those cells. Recent study in cholangiocarcinoma cells demonstrated that a type 2 to type 1 transition in apoptotic activation could be induced in TRAIL resistant cancer cells by inhibiting the hedgehog pathway (hedgehog inhibition simultaneously increase the death receptor DR4 and decrease XIAP) [20]. We performed further simulations to verify that a combined death receptor activation and XIAP inhibition based strategy works better for cancer cells having heightened sensitivity to death receptor activation but also high Bcl-2 to Bax ratio. In such a scenario, even when Bcl2 (50 fold over-expression reduced to 15 fold) and XIAP (5 fold over-expression to normal) were significantly inhibited, 5 fold over-expression of Bid and Bax was not sufficient to activate the mitochondrial pathway (through a Bid-Bax type reaction) [12]. However, depending on the cancer cell type as well as pro- and anti-apoptotic protein levels, it needs to be determined whether to induce a type 1 or a type 2 [12] or a mixed type 1-type 2 activation. Monte Carlo simulations can be carried out to simulate various possible options for inducing apoptotic death in cancer cells and determine the optimal strategy.

## 4. Conclusions

In this work, we obtain quantitative information regarding the systems level regulation of the type 1/type 2 choice in apoptotic death. Activation level of caspase 8 (initiator caspase) emerges as a key regulator of the type 1/type 2 choice and stochastic to deterministic transition in apoptotic activation. Our results indicate that increased amount of DISC clustering (presumably by higher expression of death receptors and/or lower expression of decoy receptors) results in increased caspase 8 activation (in a rapid manner). Thus cell types with high death receptor expression are expected to activate the type 1 pathway more frequently making them type 1 prone (for that specific death receptor) [5]. Conversely, low death receptor expression would make cells more prone to type 2 activation. However, over-expression of certain molecules and/or inhibition of some others may allow activation of the type 1 pathway in type 2 prone cells (or the type 2 pathway in type 1 prone cells). We also elucidate various mechanisms for mixed type 1 and type 2 activation. Expression levels of various downstream (of caspase 8) signaling molecules and kinetic reaction rates were shown to determine the time needed for completion of caspase 3 activation (through the two pathways) and impact the type 1/type 2 choice. Hence, the type 1/type 2 choice in apoptotic activation is regulated at a systems level. Interestingly, our results indicate that the type 1/type 2 choice is linked to cell-to-cell stochastic variability in apoptosis activation. We elucidate the following contributions to cell-to-cell stochastic variability: (i) time-to-caspase 8 generation (membrane proximal signaling module), (ii) time-to-cytochrome c release (pre-mitochondrial signaling module) and (iii) time-to-apoptosome formation (post-mitochondrial signaling module). The magnitude of cell-to-cell variability depends on the cellular context as well as the mechanism of activation.

In cancer therapy, it is important to perturb the apoptotic signaling network in such a manner that cancer cells are selectively targeted and cell-to-cell stochastic variability in apoptotic activation is minimized. Interestingly, these two aspects are intricately linked in apoptosis activation of cancer cells. In this study, we show that death receptor activation (such as by death ligands), along with XIAP inhibition, can lead to rapid activation of the type 1 pathway selectively in certain cancer cells. Thus it provides a mechanism for type 2 / mixed type 1-type 2 (slow and often stochastic) to type 1 (rapid and deterministic) transition in cancer cells. This chemotherapeutic strategy might work for resistant cancer cells having high Bcl2 to Bax ratio but increased sensitivity to death receptor activation. Normal cells having lower death receptor expression, in contrast, would remain protected due to slow activation and inherent variability of the type 2 pathway (under the application of death ligand and XIAP inhibitor). Our studies indicate how in silico experiments can be performed to determine the optimal strategy for targeting the apoptotic pathway of cancer cells in a cell-type specific manner.

## Appendix

The kinetic rate equations for active caspase 8 reacting with procapsase 3 (type 1 pathway) and Bid (type 2 pathway) are provided below [6, 52, 53]. [] denotes concentration of a given signaling molecule. We use the following abbreviations: C8 (active caspase 8), C3 (pro caspase 3), C3* (active caspase 3).

$$\frac{d[C8]}{dt} = -k_{on}^1[C8][C3] + k_{off}^1[C8\_C3] + k_{cat}^1[C8\_C3] - k_{on}^2[C8][Bid] + k_{off}^2[C8\_Bid] + k_{cat}^2[C8\_Bid]$$

$$\frac{d[C3]}{dt} = -k_{on}^1[C8][C3] + k_{off}^1[C8\_C3]$$

$$\frac{d[C8-C3]}{dt} = k_{on}^1[C8][C3] - k_{off}^1[C8\_C3] - k_{cat}^1[C8\_C3]$$

$$\frac{d[C3*]}{dt} = k_{cat}^1[C8\_C3]$$

$$\frac{d[Bid]}{dt} = -k_{on}^2[C8][Bid] + k_{off}^2[C8\_Bid]$$

$$\frac{d[C8-Bid]}{dt} = k_{on}^2[C8][Bid] - k_{off}^2[C8\_Bid] - k_{cat}^2[C8\_Bid]$$

$$\frac{d[tBid]}{dt} = k_{cat}^2[C8\_Bid]$$

An approximate estimation of caspase 8 – procaspase 3 and caspase 8 – Bid complexes is carried out by (i) assuming that the system has reached a steady state and (ii) ignoring the catalytic activation terms ($k_{cat}$):

$$[C8\_C3] = k_A^1[C8][C3]$$

$$[C8\_Bid] = k_A^2[C8][Bid]$$

where $k_A = k_{on} / k_{off}$.

$$\frac{[C8\_Bid]}{[C8\_C3]} = \frac{k_A^2[C8][Bid]}{k_A^1[C8][C3]} = \frac{k_A^2(Bid^0 - [C8\_Bid])}{k_A^1(C3^0 - [C8\_C3])}$$

$C3^0$ and $Bid^0$ denote the initial concentrations of procapsase 3 and Bid, respectively. For low caspase 8 concentrations (~ 1 nM), $[C8\_C3] \ll C3^0$ (100 nM) and $[C8\_Bid] \ll Bid^0$ (33 nM). In such a limit,

$$\frac{[C8_{Bid}]}{[C8_{C3}]} = \frac{k_A^2[C8][Bid]}{k_A^1[C8][C3]} = \frac{k_A^2}{k_A^1}\frac{Bid^0}{C3^0} \sim 10^3$$

## Acknowledgments

Support from NIH grant 3-CB70442 is acknowledged.

## Conflict of Interest

The authors declare no conflict of interest.

Figure S1. Single cell caspase 3 activation in Bid deficient cells for the following values of XIAP inhibition: XIAP = 60 nM, (a), XIAP = 30 nM, (b), XIAP = 0 nM (c). Smac concentration is kept constant at 50 nM. Data is shown for 10 representative single cells for each XIAP concentrations; each color corresponds to apoptosis activation for a single cell (Monte Carlo run). Decreasing XIAP concentration results in increased type 1 activation.

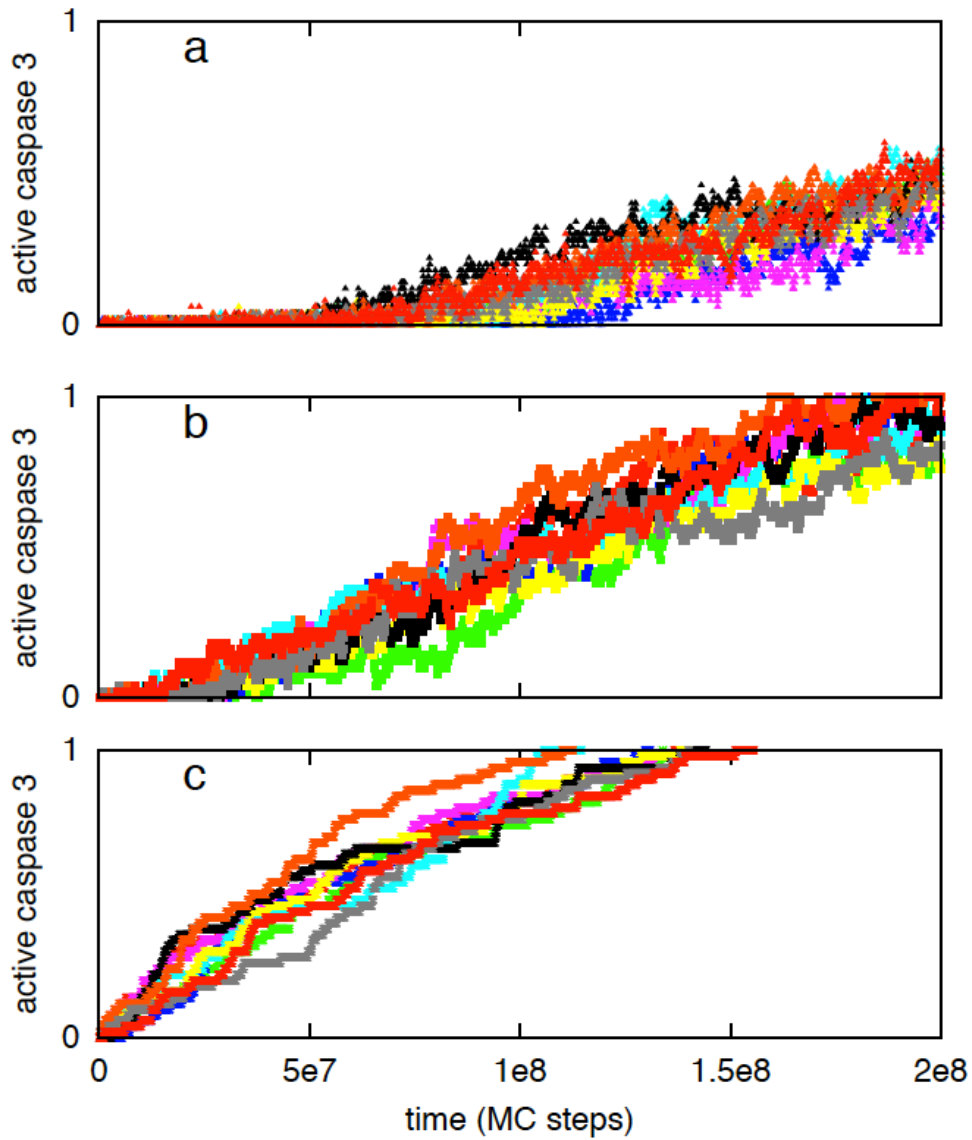

Figure S2. Single cell caspase 3 activation in cells where XIAP inhibition is removed. Data is shown for 10 representative single cells for each of the following caspase 8 concentrations = 1 nM (a), 2 nM (b) and 3 nM (c); each color corresponds to apoptosis activation for a single cell (Monte Carlo run).

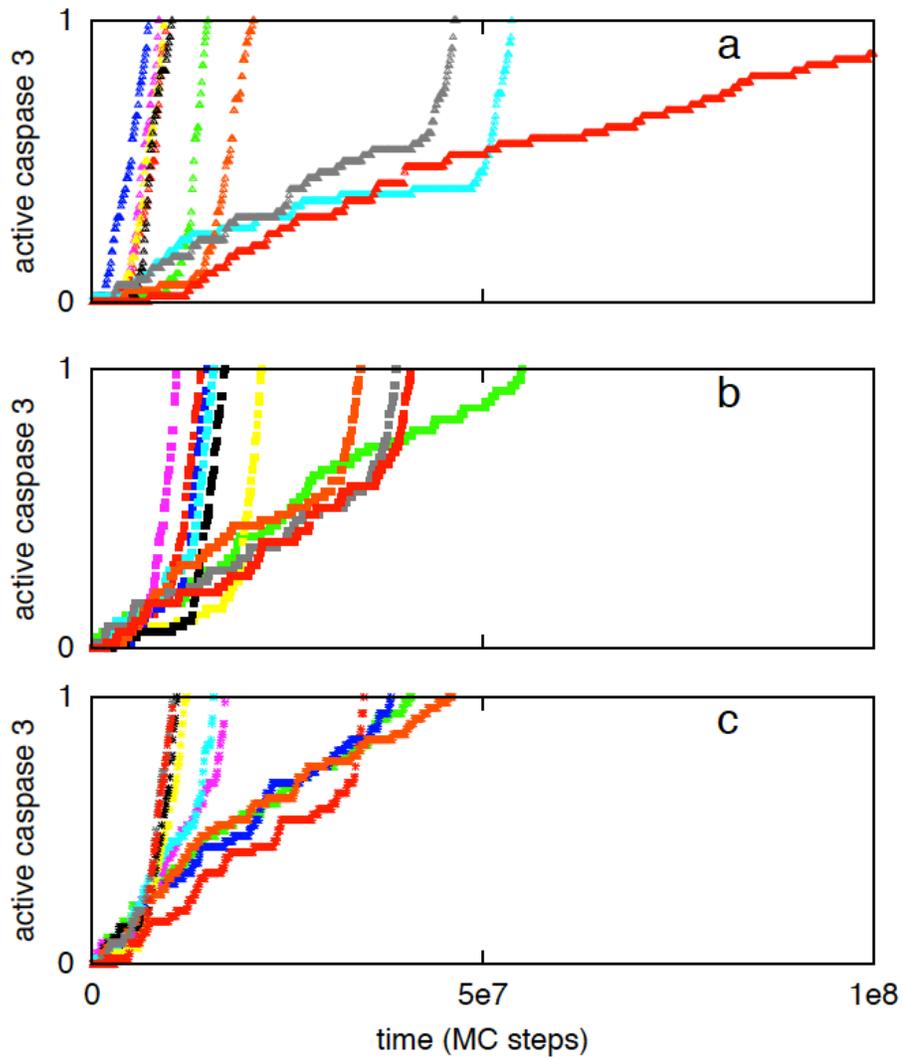

Figure S3. Fraction of apoptotic cells (measured by half-maximal caspase 3 activation) in WT and Bid[-/-] cells at different time-points. Data is obtained from 64 single cell experiments (Monte Carlo runs) with active caspase 8 ~ 5 nM. Such moderate level of caspase 8 activation might be possible to achieve in type 1 cells (at least for early times). When the type 2 pathway is blocked (by removing Bid), strong suppression of caspase 3 activation was observed for early times but late time apoptosis was not inhibited. (Results from our *in silico* studies can be compared with previous experimental data in type 1 thymocyte cells [1]: ICAD (inhibitor of caspase-activated DNase) cleavage and fluorogenic measurement of effector caspase activity (DEVDase) indicate suppression of apoptotic activation at early time (~ 1 hr) in Bid deficient thymocytes. The early-time suppression of apoptosis could be due to lack of capsase 9 activation; it is also possible that reduced apoptosis in Bid deficient cells results from lack of type 2 mediated inhibition of XIAP (by Smac/DIABLO). In our *in silico* experiments, both XIAP inhibition and caspase 9 activation contribute to early-time activation of effector caspases. Therefore, this data indicates a role of the type 2 pathway in apoptotic activation in type 1 cells such as thymocytes.)

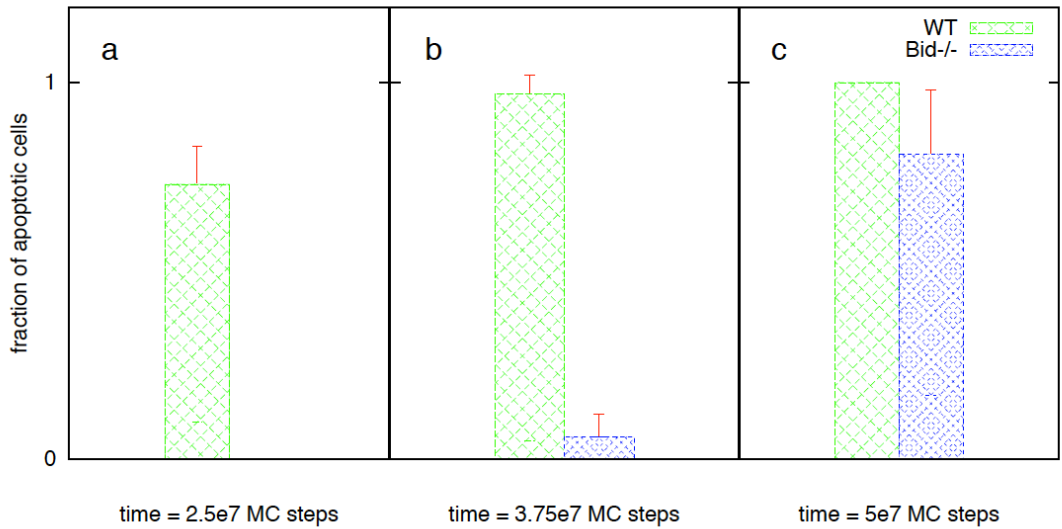

Figure S4. Single cell caspase 3 activation with concentration of Apaf reduced to 20 nM. Data is shown for 8 representative single cells; each color corresponds to apoptosis activation for a single cell (Monte Carlo run). Caspase 9 activation is slower (compared with the same for Apaf = 100 nM) due to decreased apoptosome formation and allows type 1 activation to progress longer.

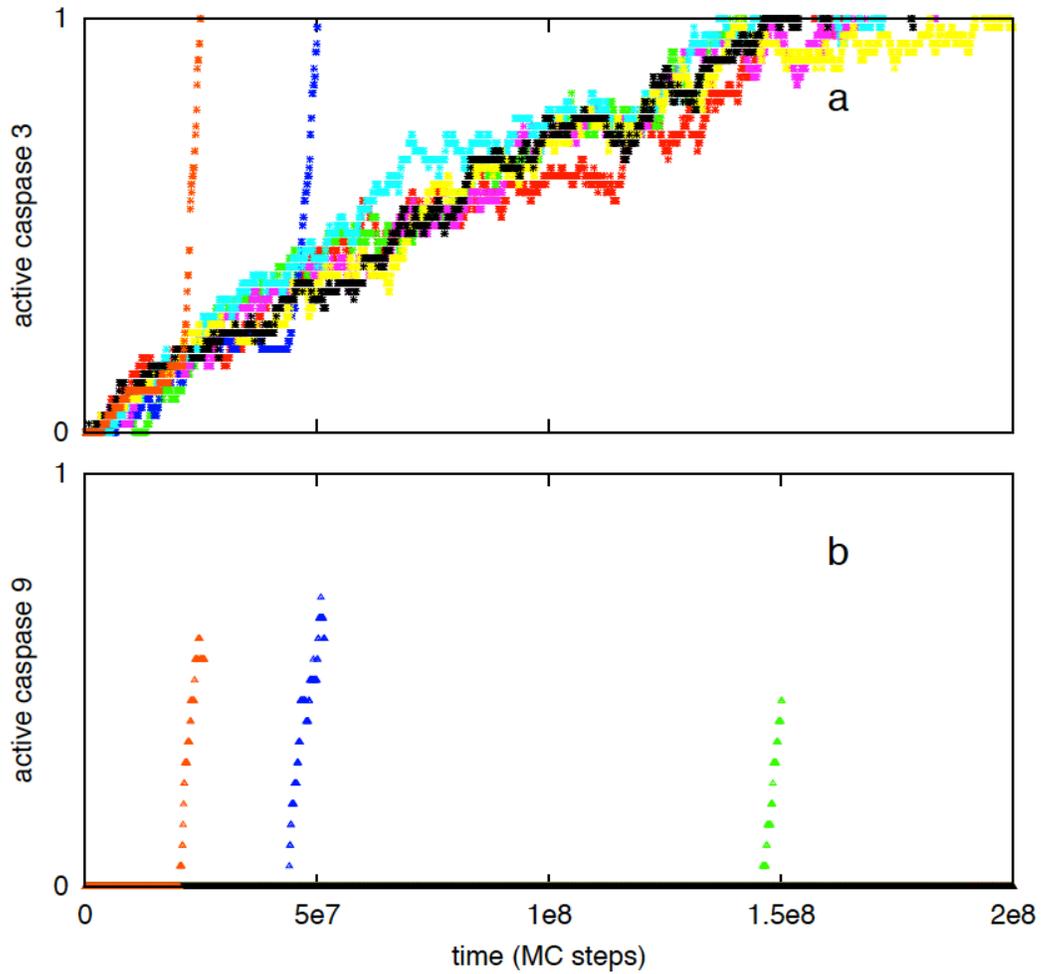

Figure S5. A previously developed model of minimal network [2] is utilized to elucidate some of the robust systems level regulatory mechanisms in apoptosis signaling. The following rate constants were used: $2 \times 10^{-4}$ (for the type 1 pathway); $1.0$, $10^{-7}$ and $10^{-1}$ (for the type 2 pathway). $X_0$ measures the strength of an apoptotic stimulus (varied in Fig S6). Initial numbers for all other the molecules ($X_1$, $X_2$, $X_3$) were assumed to be 100.

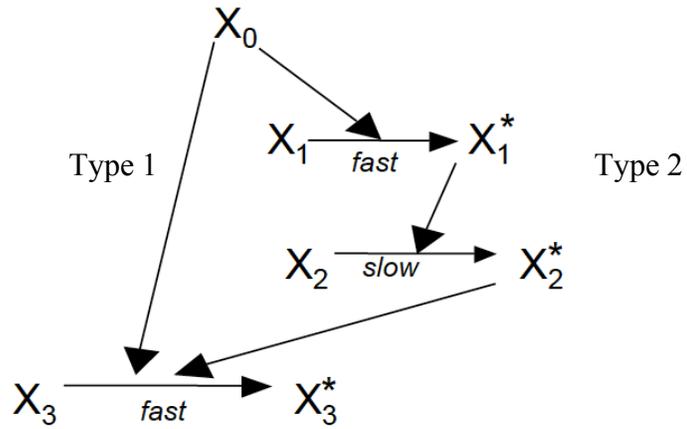

Figure S6. The effect of concentration ($X_0$) variation of the molecule that opens the type 1 – type 2 loop network. Two different values of $X_0$ were studied: 1 (left panel) and 20 (right panel). Results are shown to be robust for a 2-fold variation in concentration fluctuations for other molecules: normal concentration (a and d), 2 fold underexpression (b and e), 2 fold overexpression (c and f). The minimal network was analyzed using the stochastic differential equations developed in [2]. The SDE that captures low probability reaction of $X_2 \rightarrow X_2^*$ (slow step) was solved using a Poisson Runge-Kutta scheme [3, 4]. In this method, the number of times a specific reaction channel fires in a given time $t$ is a Poisson random variable with mean $\lambda t$ and variance $\lambda t$, where $\lambda$ is the propensity function. For the slow reaction under consideration $\lambda = k\,(\delta t)\,X_1^* X_2$. The rate constant for the slow step is $k$ and $\delta t$ is the time-step used in the numerical solution. All other reaction SDEs were solved using the standard Euler-Maruyama numerical scheme for solving stochastic differential equations [3, 5].

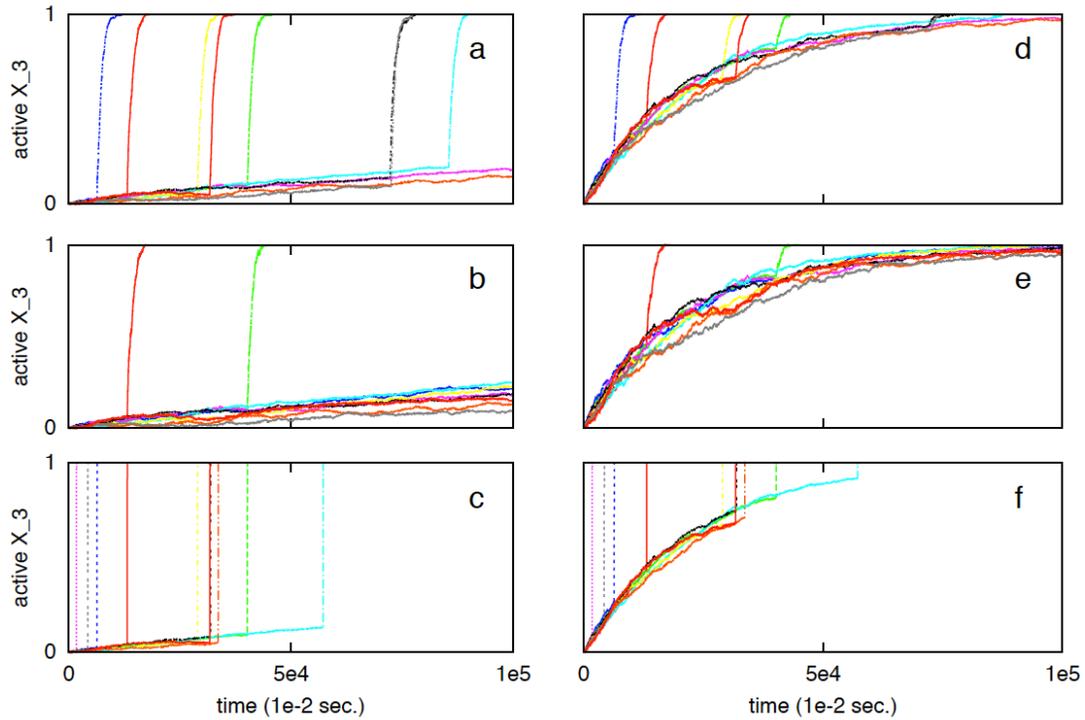

Figure S7. The effect of variation in the rate constant for the slow step ($X_2 \to X_2^*$) is studied for the following values: $10^{-6}$, $10^{-7}$ and $10^{-8}$. The minimal network was analyzed using the stochastic differential equations developed in [2]. The method utilized to solve the SDEs is the same as in Fig S6. $X_0$ (apoptotic stimulus strength) was taken as 1. (A similar study using Gillespie's stochastic simulation alogorithm (SSA) was carried out in [2]).

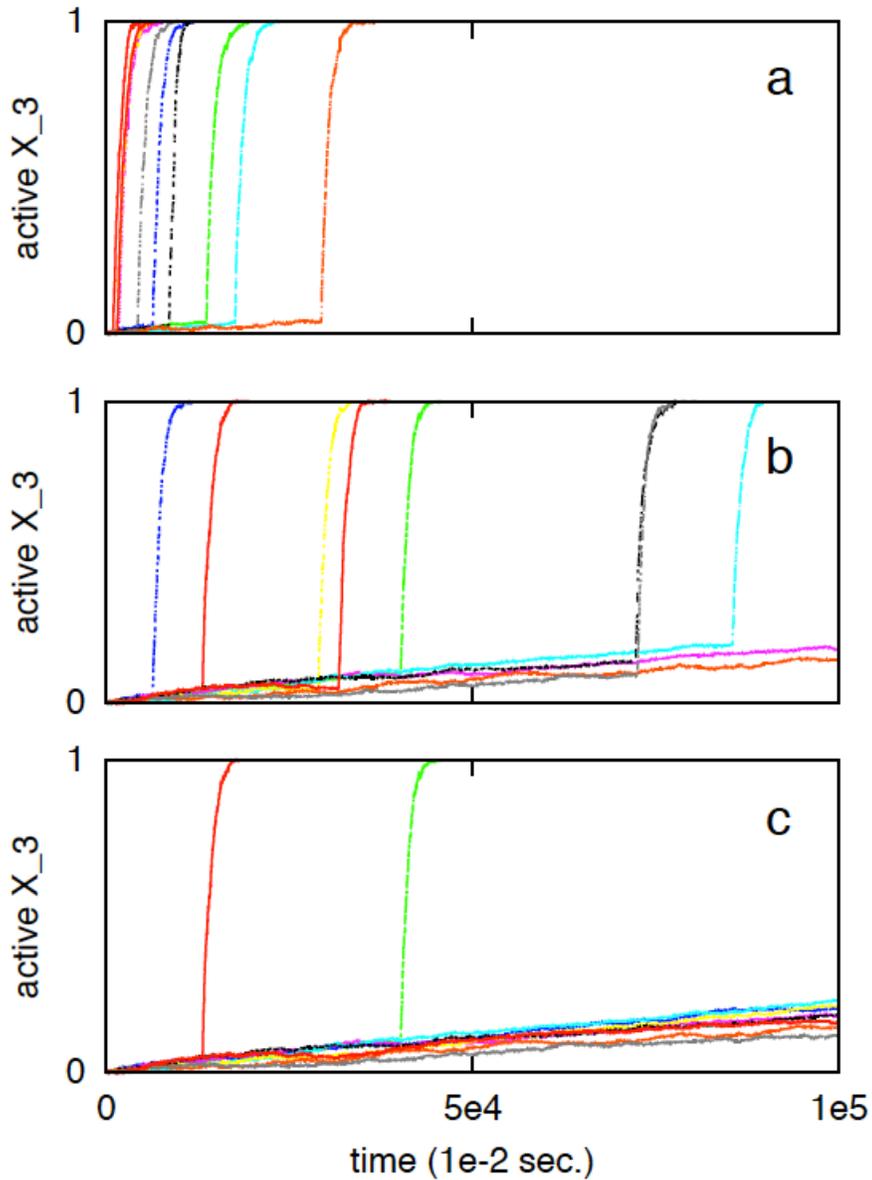

Figure S8. Time course of caspase 3 activation in cancer cells having over-expressed Bcl2 and XIAP. Bcl2 over-expression is taken as ~ 50 fold while XIAP expression is varied: normal (a and d), 3 fold (b and e), 5 fold (c and f). Such high Bcl2 expression can result from combined effect of various Bcl2 like anti-apoptotic proteins (also possibly found in cancer stem cells). Two different caspase 8 concentrations are used: 1 nM (left panel: a,b,c) and 10 nM (right panel: d,e,f). Data is shown for 5 representative single cells in each of the above cases; each color corresponds to apoptosis activation for a single cell (Monte Carlo run).

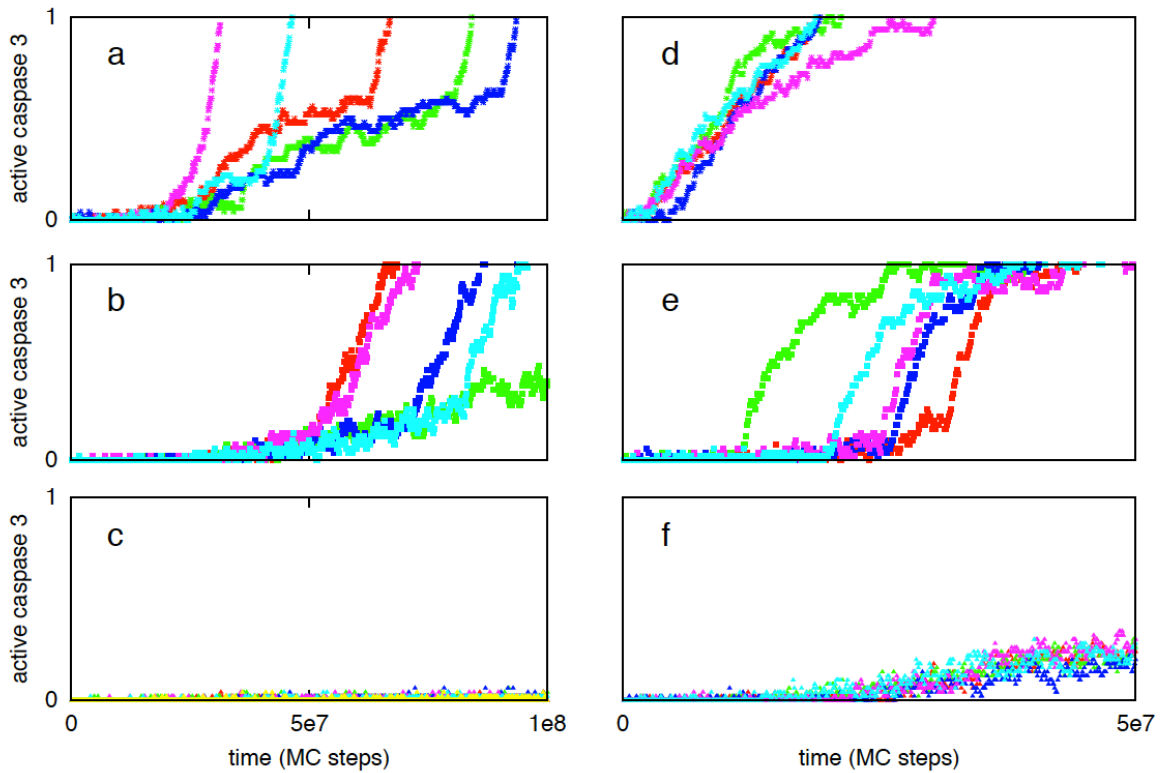

Figure S9. DISC (death inducing signaling complex) clustering for 3 different values of the DISC clustering energy parameter: $E_{dd}$ = 0, - $K_BT$ and - $2K_BT$ (a,b,c). Adapter molecules (that bind death-receptor ligand complex to generate DISC) are shown on the cell surface of 60 × 60 lattice sites (1.2 µm × 1.2 µm). Pon = 1 and Poff = $10^{-3}$ are used in the simulations. Representative single cell data are shown after $10^7$ MC steps.

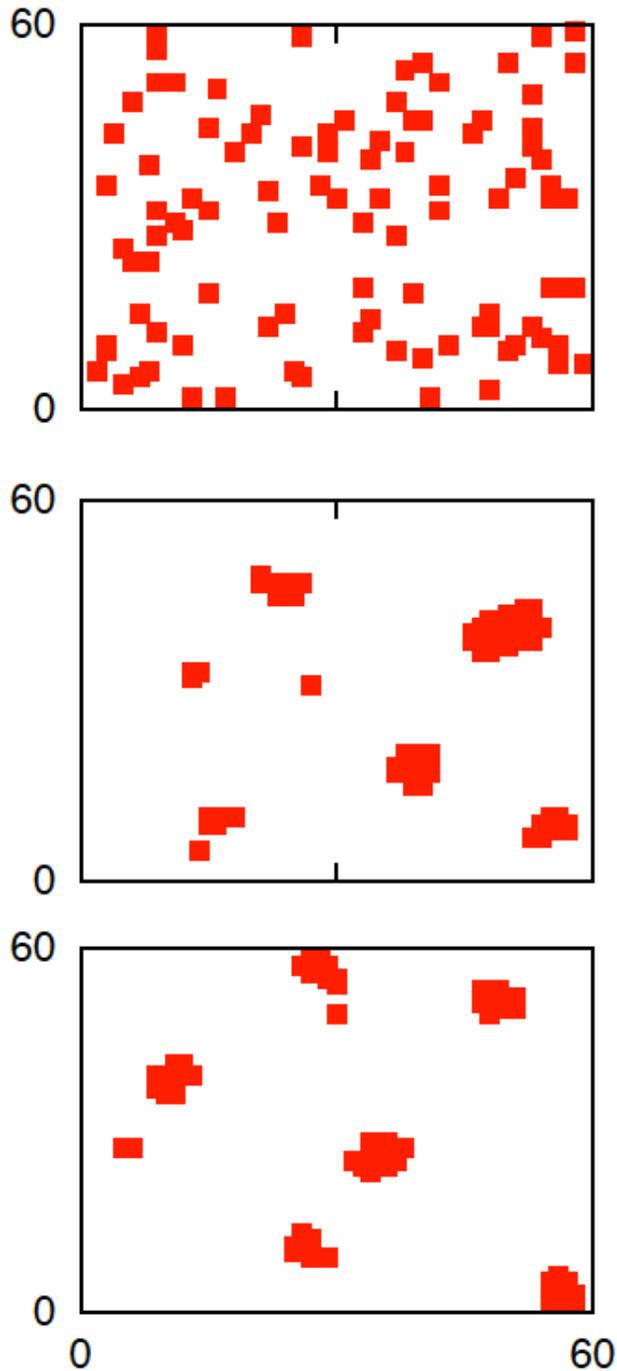

Table S1. Caspase 3 activation for different XIAP/Smac ratios and at different time-points. Active caspase 8 = 1 nM. Data normalized to half-maximal caspase 3 (= 50) activation is shown. Data is averaged over 64 single cells (MC runs) for XIAP = 30 nM and over 10 single cell runs for XIAP = 90 nM.

| XIAP/Smac combinations | Average Caspase 3 activation (SD) | | |
|---|---|---|---|
| | $T = 2.5 \times 10^7$ (MC steps) | $T = 5 \times 10^7$ (MC steps) | $T = 1 \times 10^8$ (MC steps) |
| XIAP = 30 nM Smac = 50 nM (ratio 0.6) | 0.65 (0.38) | 0.92 (0.20) | 1.0 (0.0) |
| XIAP = 30 nM Smac = 10 nM (ratio 3) | 0.47 (0.38) | 0.81 (0.32) | 0.98 (0.08) |
| XIAP = 90 nM Smac = 50 nM (ratio 1.8) | 0.01 (0.01) | 0.1 (0.08) | 1.0 (0.0) |

Table S2. Average time to first apoptosome formation as the rate constant for the cytochrome c-Apaf binding is varied. Data is averaged over 56 single cells (MC runs).

| Kinetic rate constant for Cyto c – Apaf binding | Time to (first) apoptosome formation (MC steps) |
|---|---|
| 2 fold high | $1.03 \times 10^7$ |
| normal | $2.47 \times 10^7$ |
| 2 fold low | $3.90 \times 10^7$ * |

* considering cells in which apoptosome has formed before caspase 3 activation reached its half maximal value (the actual average is higher than reported).

Table S3. Cell-to-cell stochastic variability in time-to-death (capsase 3 activation) results from cell-to-cell variability in (i) cytochrome c release (pre-mitochondrial signaling module) and (ii) apoptosome formation (post-mitochondrial signaling module). Time-to-Cytochrome c/Smac release is estimated from the initiation of XIAP-Smac complex formation. Time-to-apoptosome formation indicates the time-scale of first apoptotosome formation. Time-to-death is measured by the time by which active capsase 3 concentration reaches 50 nM (half-maximal level). Data is shown for 3 single cells (MC runs). Time is measured in MC steps.

|  | Cell-to-cell variability in caspase 3 activation | | |
| --- | --- | --- | --- |
|  | Time-to-Cyto c /Smac release | Time-to-apoptosome formation | Time-to-death |
| Cell 1 | $6.2 \times 10^6$ | $8.6 \times 10^6$ | $1.6 \times 10^7$ |
| Cell 2 | $7.6 \times 10^6$ | $9.1 \times 10^7$ | $9.5 \times 10^7$ |
| Cell 3 | $7.0 \times 10^6$ | $4.2 \times 10^7$ | $4.7 \times 10^7$ |